\documentstyle[aps,epsfig,twocolumn]{revtex}
\begin{document}
\draft
\flushbottom

\twocolumn[
\hsize\textwidth\columnwidth\hsize\csname @twocolumnfalse\endcsname

\title{Variational Mean Field approach to the Double Exchange Model}
\author{J. L. Alonso$^1$, L. A. Fern\'andez$^2$, F. Guinea$^3$,
V. Laliena$^1$ and V. Mart{\'\i}n-Mayor$^4$}
\address{
$^1$ Dep. de F{\'\i}sica Te\'orica, Facultad de Ciencias,
Universidad de Zaragoza, 50009 Zaragoza, Spain.\\
$^2$ Dep. de F{\'\i}sica Te\'orica, Facultad de CC. F{\'\i}sicas,
Universidad Complutense de Madrid, 28040 Madrid, Spain.\\
$^3$ 
Instituto de Ciencia de Materiales (CSIC). Cantoblanco,
28049 Madrid. Spain. \\
$^4$ Dip. di Fisica,
Universit\`a di Roma ``La Sapienza'', 
Ple. Aldo Moro 2, 00185 Roma and
INFN sezione di Roma, Italy.}
\date{July 25, 2000}
\maketitle 
\tightenlines
\widetext
\advance\leftskip by 57pt
\advance\rightskip by 57pt

\begin{abstract}
It has been recently shown that the double exchange Hamiltonian, with weak 
antiferromagnetic interactions, has a richer variety of first and second
order transitions than previously anticipated, and that such transitions 
are consistent with the magnetic
properties of manganites. Here we present a thorough discussion
of the variational Mean Field approach that leads to the these
results. We also show that the effect of the Berry
phase turns out to be crucial to produce first order
Paramagnetic-Ferromagnetic transitions near half filling with
transition temperatures compatible with the experimental situation.
The computation relies on two crucial facts: the use of a Mean Field
{\em ansatz} that retains the complexity of a system of electrons
with off-diagonal disorder, not fully taken into account by
the Mean Field techniques, and the small but significant
antiferromagnetic superexchange interaction between the localized spins.
\end{abstract}

\pacs{75.10.-b, %General theory and models of  magnetic ordering 
      75.30.Et  %Exchange and superexchange interactions
}
]
\narrowtext
\tightenlines

\section{Introduction}

Doped manganites show many unusual features, the most striking being
the colossal magnetoresistance (CMR) 
in the ferromagnetic (FM) phase~\cite{WK55,KS97,CVM99}.
In addition, the manganites have a rich phase diagram as function
of band filling, temperature and chemical composition. The broad
features of these phase diagrams can be understood in terms of
the double exchange model (DEM) \cite{Z51,AH55}, 
although Jahn-Teller deformations~\cite{MLS95}
and orbital degeneracy may also play a role\cite{BKK99}.
A remarkable property of these compounds is the existence of
inhomogeneities in the spin and charge distributions
in a large range of dopings, compositions
 and temperatures\cite{Tetal97,Uetal99,Fetal99}. In fact, for materials
displaying the largest CMR effects, the size of the phase-separated domains
is so large ($\sim 0.5\mu \mathrm{m}$~\cite{Uetal99}), that the electrostatic
stability of the material should be addressed by theorists. 
At band fillings where CMR effects are present, $x \sim 0.2 - 0.5$,
these compounds can be broadly classified into those with a high
Curie temperature and a metallic paramagnetic (PM) phase, and those
with lower Curie temperatures 
and an insulating magnetic phase\cite{Fetal96,Fetal98,Mira99}.

The double exchange mechanism was introduced by Zenner \cite{Z51}
through the following Kondo lattice type model
\begin{equation}
H_{\mathrm KM}=\sum_{i,j,\alpha} t_{ij}
c^\dagger_{i\alpha}c_{j\alpha}+J_{\mathrm H}\sum_{i,\alpha,\alpha'}
\mbox{\boldmath$S$}_i\cdot 
c_{i\alpha}^\dag \mbox{\boldmath$\sigma$}_{\alpha\alpha'} c_{i\alpha'}\,,
\end{equation}
where $t$ and $J_{\mathrm H}$ are, respectively, $e_{\mathrm g}$
electron's
hopping and Hund's coupling between the $e_{\mathrm g}$ and the
localized 
$t_{\mathrm 2g}$ electrons responsible for the core spin $\mbox{\boldmath$S$}$.

When $J_{\mathrm H}$ is larger than the width of the conduction band,
the model can be reduced to the double exchange model with weak
inter-atomic antiferromagnetic (AFM) interactions.  Early
investigations\cite{G60} showed a rich phase diagram, with AFM, canted
and FM phases, depending on doping and the strength of the AFM
couplings.  More recent studies have shown that the competition
between the double exchange and the AFM couplings leads to phase
separation into AFM and FM regions, suppressing the existence
of canted phases\cite{N96,RHD97,AG98,Yetal98}.  In addition, the
double exchange mechanism alone induces a change in the order of the
FM transition, which becomes of first order, and leads to phase
separation, at low doping\cite{AGG99}.  Note, however, that a
systematic study of the nature of the transition at finite temperature
was not addressed until recently~\cite{Aetal00}, despite its obvious
relevance to the experiments. In fact, in Ref.\cite{Aetal00} it was
shown that a small AFM uniform superexchange interaction between the
localized $t_{\mathrm 2g}$ spins is crucial to understand some of the
more relevant features of the phase diagram of the manganites. In
particular a first-order phase transition is found between the
PM and FM  phases in the range $x\sim 0.2-0.5$ is
found. This transition does not involve a significant change in
electronic density, so that domain formation is not suppressed by
electrostatic effects. Therefore, we find a phase-separation of a
rather different type of the previously discussed, not driven by a
charge instability, but by a {\em magnetic} instability. In addition
to this phase transition, we recover those previously discussed.

In this work we give a detailed exposition of the new mean-field
technique~\cite{Aetal00} and emphasis is made on the importance of the
Berry phase for the existence of first order phase transitions near
half filling. We have been able of achieving a more complete
description of the phase diagram than in previous work, because we
have taken full profit of a very particular feature of the DEM,
namely, fermions are bilinearly coupled to {\em classical} degrees of
freedom (the Mn spins). This allows to trace-out the fermions, thus
obtaining a non local effective Hamiltonian for the spins, that can be
explicitly written down in terms of the density of states of the
fermionic hopping matrix. What we propose to do is to calculate {\em
exactly} the effective spin-Hamiltonian, for a given (disordered in
general) configuration of the spins, using the so called {\em
moments-method}~\cite{MOMENTI} complemented with an standard
truncation procedure~\cite{TRUNCACION}. This technique can be directly
used for other models, like for instance models of classical spins and
lattice vibrations coupled to fermions without direct
interactions~\cite{MLS95}, or also in contexts different from the
manganites physics like the pyrochlores or double perovskytes.  Once
electrons are traced-out, we study the spin thermodynamics using the
variational version of the Weiss Mean-Field method~\cite{PARISI}. 

The main difference of our approach with previous work is in that we
use the exact spin-Hamiltonian, while an approximated form was used up
to now. For instance, the effective crystal approximation, which
amounts to consider that electrons move on a perfect crystal, with a
magnetically reduced hopping, was employed in~\cite{G60} (see
section~\ref{COMPARISON} for a comparison with our method).  A more
accurate estimate of the density of states, well-known from the
physics of disordered-systems and which becomes exact on the infinite
dimensions limit~\cite{AGG99}, is the CPA. Notice that for non
self-interacting electrons the Dynamical Mean-Field
Approximation(DMFA)~\cite{DYNMF,FURUKAWA} is also equivalent to a CPA
calculation on a given mean-field, which is determined via
self-consistency equations (both the mean-field and the CPA density of
states are obtained self-consistently). Although those approaches are
reasonable, and provide useful information, they are approximated in
two different ways, which is undesiderable because two different
effects are entangled. First, even for only-spins models of magnetism
(like the Ising Model) where the exact spin Hamiltonian can be
evaluated easily, the Mean-Field approximation neglects the spatial
correlations of the statistical fluctuations of the order parameter
(see e.g.~\cite{PARISI}).  But, in addition, for an electron-mediated
magnetic interaction, the evaluation of the effective spin Hamiltonian
is only accurate in the limit of infinite dimensions. With our
approach, the correlations on the magnetic fluctuations are to be
blamed for all the differences between our results and the real
behavior of the model. On the other hand, in section~\ref{COMPARISON}
we show how the failure of the effective-crystal approximation on
finding the first-order phase transition at half-filling is due to the
inaccuracy on the calculation of the density of states.  Moreover, we
are able to study directly in three dimensions some rather subtle
details, like the non-negligible effects of keeping the Berry phase on
the DEM Hamiltonian.

The structure of the paper is as follows. In Section II we introduce the
DEM and our notations. In Section III we present our Mean-Field 
approximation. The very non-trivial part of the work, the computation
of the Density of States, is explained in Section IV. It requires numerical 
simulations that can be performed on large lattices with a high accuracy.
They are described in Section V. 
The effects of the Berry phase are analyzed in Section VI.
Section VII is devoted to the study of the influence of the Berry phase in
the phase diagram of the DEM. The comparison of
the Mean-Field approach studied in this work with the de Gennes's
\cite{G60} and with the DMFA is carried out in Section VIII.
The conclusions are summarized in Section~IX. 

\section{Model}

We study a cubic lattice with one orbital per site. At each site there is
also a classical spin. The coupling between the conduction electron and this 
spin is assumed to be infinite, so that the electronic state with spin
antiparallel to the core spin can be neglected. Finally,
we include an AFM coupling between nearest neighbor core
spins. We neglect the degeneracy of
the conduction band. Thus, we cannot analyze effects related
to orbital ordering, which can be important in the highly doped
regime, $x > 0.5$\cite{BKK99,Hetal00} (see however Ref.~\cite{Khom00}).
We also neglect the
coupling to the lattice. We focus on the role of the magnetic interactions
only. As mentioned below, magnetic couplings
suffice to describe a number of discontinuous transitions
in the regime where CMR effects are observed.
These transitions modify substantially the coupling between
the conduction electrons and the magnetic excitations.
Thus, they offer a simple explanation for the anomalous
transport properties of these compounds. Couplings to additional
modes, like optical or acoustical phonons, will enhance further the
tendency towards first order phase transitions. We consider that
a detailed understanding of the role of the magnetic interactions
is required before adding more complexity to the model. 
Note that there is indeed evidence that, in some compounds,
the coupling to acoustical phonons\cite{Ietal95} or to Jahn-Teller
distortions\cite{Betal98} is large.

The Hamiltonian of the DEM is

\begin{equation}
{\cal H} = \sum_{ij}
 {\cal T}( \mbox{\boldmath$S$}_i,\mbox{\boldmath$S$}_j )
c_i^\dag c_j +
\sum_{ij}\tilde{J}_{\mathrm AF} 
S^2\mbox{\boldmath$S$}_i\cdot\mbox{\boldmath$S$}_j
\label{hamil}
\end{equation}

\noindent
where $S = 3/2$ is the value of the spin of the core, Mn$^{3+}$, and
$\mbox{\boldmath$S$}$ stands for a unit vector oriented parallel to
the core spin, which we assume to be classical. In the following, we
will use $J_{\mathrm AF} = \tilde{J}_{\mathrm AF} S^2$. Calculations
show that the quantum nature of the core spins does not induce
significant effects\cite{AG98}. In one of the earliest studies of this
model \cite{G60}, the superexchange coupling was chosen FM between
spins lying on the same $z=constant$ plane, and AFM between spins
located on neighboring planes. This is a reasonable starting point for
the study of ${\mathrm La}_{1-x}{\mathrm Ca}_x{\mathrm Mn}{\mathrm
O}_3$ if $x<0.16$, where A-type antiferromagnetism is found. For larger
doping, $0.16<x<0.5$, which is our main focus, the magnetism is
uniform and there is no a priori reason for favoring a particular
direction.

The function

\begin{equation}
{\cal T}(\mbox{\boldmath$S$}_i,\mbox{\boldmath$S$}_j )
=t\left[
\cos\frac{\theta_i}{2}\cos\frac{\theta_j}{2}+
\sin\frac{\theta_i}{2}\sin\frac{\theta_j}{2}
{\mathrm e}^{{\mathrm i}(\varphi_i-\varphi_j)}\right]
\label{hopping}
\end{equation}

\noindent
stands for the overlap of
two spin 1/2 spinors oriented along the directions defined by
$\mbox{\boldmath$S$}_i$ and $\mbox{\boldmath$S$}_j$, whose polar and 
azimuthal angles are denoted by $\theta$ and $\varphi$, respectively. 
It defines a hopping matrix ${\cal T}$, whose matrix elements are
${\cal T}_{ij}=
{\cal T}(\mbox{\boldmath$S$}_i,\mbox{\boldmath$S$}_j )$. The hopping
function can be written as

\begin{equation}
{\cal T}(\mbox{\boldmath$S$}_i,\mbox{\boldmath$S$}_j )=t
\cos\frac{\vartheta_{ij}}{2}\exp({\mathrm i}\phi_{ij}) \, ,
\end{equation}

\noindent
where $\vartheta_{ij}$ is the relative angle between 
$\mbox{\boldmath$S$}_i$ and $\mbox{\boldmath$S$}_j$, and
$\phi_{ij}$ is the so called Berry phase. It is sometimes
assumed that the Berry phase can be set to zero without
essential loss. It is therefore interesting to study the
model that ignores the Berry phase, the hopping matrix 
being

\begin{equation}
{\cal T}^{\mathrm mod}_{ij}=
\left|{\cal T}_{ij}\right|=
\cos\frac{\vartheta_{ij}}{2}=
\sqrt{\frac{1+\mbox{\boldmath$S$}_i\cdot
\mbox{\boldmath$S$}_j}{2}} \, .
\label{hopmod}
\end{equation}

\noindent
In the following sections we will analyze both models,
with Berry phase (hopping matrix ${\cal T}$) and
without Berry phase (hopping matrix 
${\cal T}^{\mathrm mod}$).

\section{Mean-field approximation}

Our approach to the problem follows the variational formulation of 
the Mean-Field approximation, described for instance in 
\cite{PARISI}. 
We start by writing the Grand Canonical partition function for the
DEM:

\begin{equation}
{\cal Z}_{\mathrm GC} = \int[dS]{\mathrm Tr}^{\mathrm (Fock)}
\exp\left[-({\cal H}-\mu {\cal N})/T\right]\, ,
\label{partfun}
\end{equation}

\noindent
where $\mu$ is the electronic chemical potential, 
${\cal N}=\sum_i c_i^\dagger c_i$ is the electron number operator,
$T$ is the temperature and we use units in which the Boltzmann constant 
is one. The trace, taken over the electron Fock space, defines
an effective Hamiltonian for the spins,

\begin{equation}
\exp\left[-{\cal H}^{\mathrm eff}(\mbox{\boldmath$S$})/T\right] 
= 
{\mathrm Tr}^{\mathrm (Fock)}\exp\left[-({\cal H}-\mu {\cal N})/T\right] \, ,
\label{heff_def}
\end{equation}

\noindent
that can be computed in terms of the eigenvalues, $E_n$, of the 
hopping matrix, ${\cal T}$: 

\begin{eqnarray}
{\cal H}^{\mathrm eff}(\mbox{\boldmath$S$})&=&
J_{\mathrm AF}\sum_{\langle ij\rangle}  
\mbox{\boldmath$S$}_i\cdot\mbox{\boldmath$S$}_j - \nonumber \\
&&T\sum_n
\log\left\{1+
\exp\left[-(E_n(\mbox{\boldmath$S$})-\mu)/T\right]\right\}
\,.\label{HEFF}
\end{eqnarray}

\noindent
Introducing the Density of States (DOS) of
${\cal T}$:

\begin{equation}
g(E;\mbox{\boldmath$S$})=
\frac{1}{V}\sum_{n=1}^V
\delta\left[E-E_n(\mbox{\boldmath$S$})\right] \, ,
\label{dos}
\end{equation}

\noindent
where $V$ is the volume of the lattice,
the effective Hamiltonian can be written as

\begin{eqnarray}
{\cal H}^{\mathrm eff}(\mbox{\boldmath$S$})&=&
\sum_{\langle ij\rangle}J_{\mathrm AF}
\mbox{\boldmath$S$}_i\cdot\mbox{\boldmath$S$}_j-\nonumber\\
&&TV\int dE\,g(E;\mbox{\boldmath$S$})
\log\left[1+e^{-(E-\mu)/T}\right]\,.\label{heff}
\end{eqnarray}

\noindent
The Grand Canonical partition 
function becomes an integral in spin configuration space:

\begin{equation}
{\cal Z}_{\mathrm GC}=\int[dS]
\exp\left[-{\cal H}_{\mathrm eff}
(\mbox{\boldmath$S$})/T\right] \, .
\label{partfun_eff}
\end{equation}

Thermodynamics follows from Eq.~(\ref{partfun_eff}) as
usual. The free energy, ${\cal F}$, and the electron density, $x$,
are given by:
\begin{equation}
{\cal F} = -\frac{T}{V}\log{\cal Z}_{\mathrm GC} 
\end{equation}
\begin{equation}
x = \frac{\partial {\cal F}}{\partial \mu}=
\int dE\,\langle g(E;\mbox{\boldmath$S$})\rangle
\frac{1}{1+\exp\left[(E-\mu)/T\right]} 
\end{equation}

\noindent
where $\langle\cdots\rangle$ stands for spectation value over equilibrium
spin configurations.

The variational Mean-Field approach consist on comparing 
the actual 
system with a set of simpler reference models, whose 
Hamiltonians, ${\cal H}_{\mbox{\scriptsize\boldmath$h$}}$,
depend on external parameters, $\mbox{\boldmath$h$}_i$. 
For simplicity, we choose the model:
\begin{equation}
{\cal H}_{\mbox{\scriptsize\boldmath$h$}}=-\sum_i
\mbox{\boldmath$h$}_i\cdot\mbox{\boldmath$S$}_i
\label{hamil_mf} \, .
\end{equation}
The variational method is based on the inequality

\begin{equation}
{\cal F}\leq{\cal F}_{\mbox{\scriptsize\boldmath$h$}}+
\left\langle{\cal H}^{\mathrm eff}
-{\cal H}_{\mbox{\scriptsize\boldmath$h$}}
\right\rangle_{\mbox{\scriptsize\boldmath$h$}} 
 , \label{ineq}
\end{equation}

\noindent
where ${\cal F}_{\mbox{\scriptsize\boldmath$h$}}$ is the free energy of 
the system 
with Hamiltonian (\ref{hamil_mf}), and the expectation values 
$\langle\cdots\rangle_{\mbox{\scriptsize\boldmath$h$}}$
are calculated with the Hamiltonian 
${\cal H}_{\mbox{\scriptsize\boldmath$h$}}$. 
The inequality (\ref{ineq}) follows easily from the concavity 
of the exponential function \cite{PARISI}.
The best approximation to the actual free energy with the 
{\em ansatz} of Eq.~(\ref{hamil_mf}) is

\begin{equation}
{\cal F}=\min_{\mbox{\scriptsize\boldmath$h$}}
\left\{{\cal F}_{\mbox{\scriptsize\boldmath$h$}}+
\left\langle{\cal H}^{\mathrm eff}
\right\rangle_{\mbox{\scriptsize\boldmath$h$}} 
-\left\langle{\cal H}_{\mbox{\scriptsize\boldmath$h$}}
\right\rangle_{\mbox{\scriptsize\boldmath$h$}}
\right\} \, .
\label{mfeq}
\end{equation}

Since, for technical reasons that will become clear in the following,
it is not possible to work with one field $\mbox{\boldmath$h$}_i$
per site, we must select some subsets that contain only a few
independent parameters (see Section VII). The choice is of paramount 
importance since
it is an {\em ansatz} that will artificially restrict the behavior
of the system. We have chosen the following four families \cite{remark}
of fields, depending on a parameter, $\mbox{\boldmath$h$}$:
\begin{eqnarray}
\mbox{\boldmath$h$}_i&=&\mbox{\boldmath$h$}\,,\\
\mbox{\boldmath$h$}_i&=&(-1)^{z_i}\mbox{\boldmath$h$}, \\
\mbox{\boldmath$h$}_i&=&(-1)^{x_i+y_i}\mbox{\boldmath$h$}, \\
\mbox{\boldmath$h$}_i&=&(-1)^{x_i+y_i+z_i}\mbox{\boldmath$h$}\,,
\end{eqnarray}
which correspond, respectively, to FM, A-AFM, C-AFM, and G-AFM orderings.
There is an order parameter (magnetization) associated to each of these 
orderings. We will denote them by $M_{\mathrm F}$, $M_{\mathrm A}$, 
$M_{\mathrm C}$, and $M_{\mathrm G}$, 
respectively. As a shorthand, they will be denoted generically by
${\cal M}$. The order parameter 
is related to the corresponding $\mbox{\boldmath$h$}$ by
\begin{equation}
{\cal M}=\frac{1}{\tanh h}-\frac{1}{h}\,,\label{tanh}
\end{equation}
where $h=|\mbox{\boldmath$h$}|/T$. Thus, the free energy can be
written in terms of ${\cal M}$ instead of $h$ and Eq.~(\ref{mfeq}) 
implies that it must be minimized  with respect
to $\cal M$. The free energy has three contributions: the fermion free
energy (FFE), the superexchange energy and the entropy of the spins:
\begin{equation}
{\cal F}({\cal M})={\cal F}_{\mathrm Fer}({\cal M})+N
J_{\mathrm{AF}}{\cal M}^2-
T{\cal S}_{\mbox{\scriptsize\boldmath$h$}}({\cal M})\,,
\label{entropy}
\end{equation}
where $N$ is , respectively, 3,$-3$, 1 and $-1$ for FM, G-AFM, A-AFM and
C-AFM orderings. 

The entropy
of the spins can be easily computed in terms of the
Mean-Field: 
\begin{eqnarray}
{\cal S}_{\mbox{\scriptsize\boldmath$h$}}({\cal M}) &=&(
{\cal F}_{\mbox{\scriptsize\boldmath$h$}} - \left\langle
{\cal H}_{\mbox{\scriptsize\boldmath$h$}}
\right\rangle_{\mbox{\scriptsize\boldmath$h$}})/T \nonumber \\
&=&
\log\left[\sinh h({\cal M})/h({\cal M})\right]-
h({\cal M}){\cal M} \, ,
\label{entro}
\end{eqnarray}
\noindent
but the FFE,

\begin{equation}
{\cal F}_{\mathrm Fer}({\cal M})=-T\int dE\, 
\left\langle g(E;\mbox{\boldmath$S$})
\right\rangle_{\mbox{\scriptsize\boldmath$h$}}
\log\left[1+{\mathrm e}^{-(E-\mu)/T}\right] \, ,
\label{elibre_f}
\end{equation}
must be estimated numerically.

The non trivial part of the computation is the
average of the DOS $\langle g(E;\mbox{\boldmath$S$})
\rangle_{\mbox{\scriptsize\boldmath$h$}}$. 
The key is that it can be computed by numerical simulations 
on large lattices with high accuracy, ought to two basic facts:
1) the Mean Field Hamiltonian (\ref{hamil_mf}) describes uncorrelated 
spins and
therefore equilibrium spin configurations can be easily generated 
on very large lattices, and 2) the DOS is a self-averaging
quantity. This last point means that the mean value of the DOS 
can be obtained on a large lattice by averaging it over a small set
of equilibrium configurations. Once the DOS is
computed, the integral of Eq.~(\ref{elibre_f}) can
be performed numerically to get the FFE
as a function of ${\cal M}$.

\section{Computation of the averaged DOS}

The DOS can be accurately computed for {\em any} given spin 
configuration
with the technique that we describe in the following~\cite{MOMENTI}. 
From its definition, Eq.~(\ref{dos}), 
the DOS is a probability distribution in the variable $E$,
whose moments are

\begin{equation}
\mu_k(\mbox{\boldmath$S$})=
\int dE\,g(E;\mbox{\boldmath$S$})E^k=
\frac{1}{V}{\mathrm Tr}{\cal T}^k
\label{EQMOMENTOS}
\end{equation}

\noindent
Now, it is easy to show that the eigenvalues of the hopping matrix
verify $-6t\le E_n< 6t$.  A probability distribution of compact
support can be reconstructed from its moments using the techniques of
Stieljes~\cite{CHIHARA}.  In practice, we only know the first $p$
moments, but the method of Stieljes allows us to find a good
approximation to the distribution if $p$ is large enough.

To compute the averaged DOS we follow four steps:

\begin{itemize}
\item[i)] 
Generate spin configurations, $\left\{\mbox{\boldmath$S$}\right\}$, 
according to the Mean-Field Boltzmann weight,
$\exp(-{\cal H}_{\mbox{\scriptsize\boldmath$h$}}/T)$. This can be achieved
very efficiently with a heat bath algorithm, since all the spins
are decoupled in the Mean-Field Hamiltonian. In this way, one
obtains spin configurations in perfect thermal equilibrium with the
Boltzmann-weight given by the Mean-Field Hamiltonian. 
\item[ii)] For each $\left\{\mbox{\boldmath$S$}\right\}$, one would
calculate the moments of the DOS, using Eq.~(\ref{EQMOMENTOS}), and
then apply the techniques of reconstruction of Stieljes~\cite{CHIHARA}. 
However, this is impractical since,
although the matrix ${\cal T}$ is sparse, the trace in
Eq.~(\ref{EQMOMENTOS}) would require to repeat the process $V$ times,
and we would end-up with an algorithm of order $V^2$. We use instead
an stochastic estimator. First, we extract a normalized random vector
$|v\rangle$, with components
\begin{equation}
v_i =\frac{\alpha_i}{\sum_{j=0}^V \alpha_j^2}\, 
\end{equation}
where the $\alpha_i$ are random numbers extracted with  uniform probability
between $-1$ and $1$. Let us now call $|n\rangle$ to the eigenvector of 
eigenvalue $E_n$ of the matrix ${\cal T}$. It is easy to check that
(the overline stands for the average on the random numbers $\alpha_i$)
\begin{eqnarray}
\overline{v_iv_j}&=&\frac{\delta_{i,j}}{V}\,,\\
\overline{\langle n|v\rangle\langle v|m\rangle}&=&\frac{\delta_{n,m}}{V}\,.
\label{ELPROMEDIO}
\end{eqnarray}
Then we introduce a $v$-dependent density of states:
\begin{equation}
g(E;v;\mbox{\boldmath$S$})\equiv \sum_{n=1}^V\,|\langle n|v\rangle|^2\,\delta(E-E_n)\,.
\label{GDEVDEE}
\end{equation}
From Eq.~(\ref{ELPROMEDIO}), it follows immediately that
\begin{equation}
\overline{g(E;v;\mbox{\boldmath$S$})}=g(E;\mbox{\boldmath$S$})\,.
\end{equation}
From $\langle v|v\rangle=1$, and from Eq.~(\ref{GDEVDEE}), we see
that $g(E;v;\mbox{\boldmath$S$})$ is a perfectly reasonable distribution function,
whose moments are
\begin{equation}
\int\, dE\ E^k\,g(E;v;\mbox{\boldmath$S$}) = \langle v|{\cal T}^k|v\rangle
\label{RANDOMVECTOR}
\end{equation} 
Numerically, the algorithm is of order $k\times V$, since, as
mentioned, ${\cal T}$ is sparse and only $O(V)$ operations are
required to multiply $v$ by ${\cal T}$. This method allows to compute
a large number of moments on large lattices. However, notice that the
actual calculation is not performed this way (round-off errors would
grow enormously with the power of ${\cal T}$), but as explained in the
appendix.
\item[iii)]
Reconstruct $g(E;v;\mbox{\boldmath$S$})$ from the moments, by the
method of Stieljes. The DOS is obtained in a discrete but very
large number of energies, $E$. The cost of refining this set of
energies is negligible.
Hence, the integral over $E$ that gives the FFE, 
Eq.~(\ref{elibre_f}), can be approximated numerically with 
high accuracy. 
\item[iv)] Average $g(E;v;\mbox{\boldmath$S$})$ over the spin
configurations $\mbox{\boldmath$S$}$ and over the random vectors
$|v\rangle$.  In practice we only use a random vector per spin
configuration: it is useless to obtain an enormous accuracy on
the density of states for a particular spin configuration, that should
be spin-averaged, anyway. Since the
errors due to the fluctuations of the spins and the fluctuations of
the $|v\rangle$ are statistically uncorrelated, both of them average
out simultaneously~\cite{ERRORS}.  
It is crucial that the DOS is a self-averaging
quantity, what means that its fluctuations are suppressed as
$1/\sqrt{V}$. Hence, its average over a few equilibrium configurations
is enough to estimate it with high accuracy.
\end{itemize}

This program can be carried out successfully on lattices as large 
as $64^3$ and even $96^3$, computing a large enough number of moments,
$50$ or $75$. As we shall see, this suffices to achieve
an excellent accuracy in the averaged DOS and in the FFE.

The whole process is repeated for several values of the Mean Fields
$\mbox{\boldmath$h$}$ within each family. 
The FFE computed in a discrete set of magnetizations
is extended to a continuous function of ${\cal M}$ in 
the interval $[0,1]$ through a polynomial fit, as we will show in 
next section. The magnetic phase diagram of the model will come
out easily then.

\section{Extracting the Fermionic Free Energy}

Let us discuss in this section the numerical results for the averaged
DOS and for the FFE. We carried out the program designed in the previous 
section for 20 values of $\mbox{\boldmath$h$}$ within each family
of Mean Fields, chosen in such a way that the corresponding 
{\em magnetizations}, $\cal M$, are uniformly distributed in
$[0,1]$. The expectation value $\langle g(E;\mbox{\boldmath$S$})\, 
\rangle_{\mbox{\scriptsize\boldmath$h$}}$ is
estimated by averaging over 50 equilibrium (with respect to the Mean Field
distribution) spin configurations. This is enough to have the statistical 
errors under control on a $64^3$ lattice.

\vbox{
\begin{figure}[t!]
\centerline{\epsfig{file=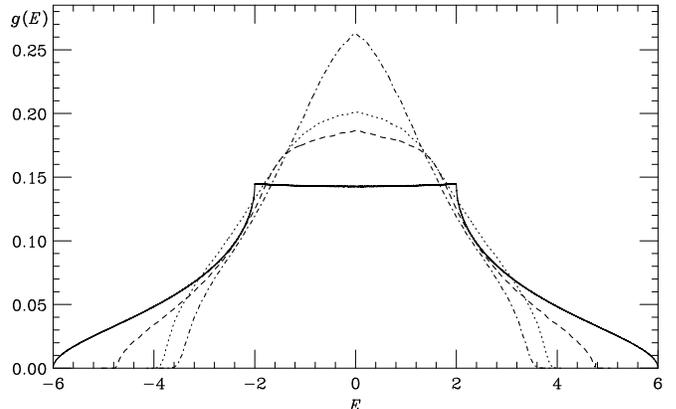,width=55mm,angle=90}} 
\caption{Averaged DOS, reconstructed with 50 moments, 
versus $E$, for four values of the mean-field
corresponding to $M_{\mathrm F}=1$ (solid line), 
$M_{\mathrm F}=0.5$, (dashed line),
paramagnetic (dotted line), and
$M_{\mathrm G}=0.5$, (dot-dashed line).}
\label{fig:dos}
\end{figure}
}

Figure~\ref{fig:dos} displays the averaged DOS, computed on a
$64^3$ lattice for four values of $\mbox{\boldmath$h$}$,
corresponding to FM, ($M_{\mathrm F}=0.5$ and $M_{\mathrm F}=1$), 
paramagnetic, and G-AFM ($M_{\mathrm G}=0.5$) phases. They were
reconstructed with its 50 first moments~\cite{ACLARACIONMOM}.
Note that the DOS is even in $E$, as required by the
the particle-hole symmetry, and that it becomes narrower in 
going from the FM to the G-AFM phase, as expected. The width of
the density of states in the PM phase is $8 t$,
two thirds of the width corresponding to the perfect ferromagnetic,
in agreement with the results of full diagonalization~\cite{CVB98}.

From the averaged DOS it is straightforward to compute the FFE by
performing the integral entering Eq.~(\ref{elibre_f}) numerically. In
this way, we computed ${\cal F}_{\mathrm Fer}$ for the chosen values
of ${\cal M}$. Given that the free energy can be shifted by a term
independent of ${\cal M}$, 
we use ${\cal F}_{\mathrm Fer}({\cal M})-{\cal F}_{\mathrm Fer}(0)$
instead of ${\cal F}_{\mathrm Fer}({\cal M})$~\cite{remark2}. 

To have an analytic expression for ${\cal F}_{\mathrm Fer}$, 
we fit the data with a polynomial of order sixth in ${\cal M}$ , 
with coefficients that depend on $T$ and $\mu$:

\begin{equation}
{\cal F}_{\mathrm Fer}^{(I)}(M_I) \;=\; 
A_2^{(I)}\,M_I^2\:+\:
A_4^{(I)}\,M_I^4\:+\:A_6^{(I)}\,M_I^6 \, .
\label{fits}
\end{equation}
The index $I$ denotes the type of ordering: $I\,=\, {\mathrm F, A, C}$, or 
${\mathrm G}$.

Figure~\ref{fig:elibre_f}
shows the FFE at $T=0$ and $\mu=0$, which
correspond to half filling, $x=1/2$, as a function of $M_{\mathrm F}$ or
$M_{\mathrm G}$. 
The points are the result of the numerical computation and the
lines are the best fits of the form (\ref{fits}).
The high quality of the fits is remarkable.
Note that the fermions favor FM order. 
The coefficients $A_j^{(I)}(T,\mu)$ of Eq.~(\ref{fits}), which 
will play 
a major role in the exploration of the phase diagram, are
displayed in Figure~\ref{fig:coeff_errors} in the cases
$I={\mathrm F,G}$, for $T=0$, as a function of $\mu$.
We always found $A_2^{\mathrm (F)}(T,\mu)<0$ and 
$A_2^{\mathrm (G)}(T,\mu)>0$, in 
agreement with the FM nature of the spin interaction induced by the double 
exchange mechanism. 

\vbox{
\begin{figure}[t!]
\centerline{\epsfig{file=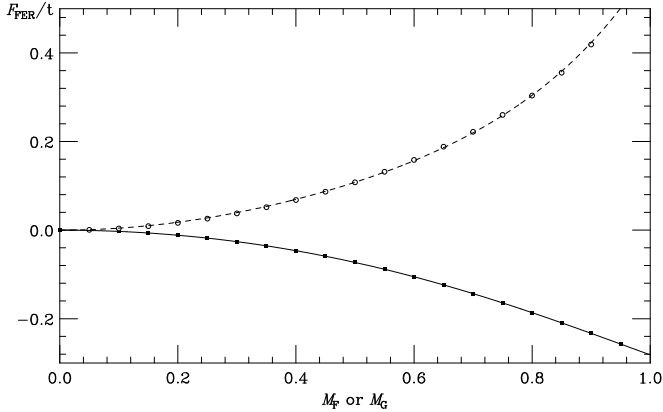,width=55mm,
angle=90}} 
\caption{FFE at $T=0$ and $\mu=0$ ($x=1/2$) versus $M_{\mathrm F}$ 
(squares, solid line) 
or $M_{\mathrm G}$ (circles, dotted line). The points are the results 
of the simulation and
the lines are the three parameter fit.}
\label{fig:elibre_f}
\end{figure}
}

\vbox{
\begin{figure}[t!]
\centerline{\epsfig{file=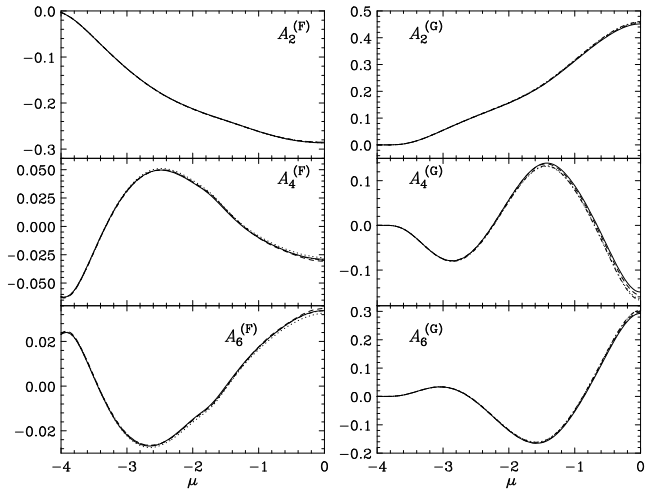,width=65mm,
angle=90}} 
\vspace*{0truecm}
\caption{Coefficients $A_2^{\mathrm (F)}$, $A_2^{\mathrm (G)}$, 
$A_4^{\mathrm (F)}$, 
$A_4^{\mathrm (G)}$,
$A_6^{\mathrm (F)}$, and $A_6^{\mathrm (G)}$ of the fits (\protect\ref{fits}) 
versus $\mu$, at $T=0$.
Each figure contains four lines (that sometimes cannot be
distinguished) corresponding to the four different computations
mentioned in the text, three
with 50 and one with 75 moments.}
\label{fig:coeff_errors}
\end{figure}
}

Let us estimate the errors of our numerical approach.
We have three sources of errors:

\begin{itemize}
\item[a)]Finite size of the lattice.
\item[b)]Statistics, arising from the numerical simulation.
\item[c)]Truncation of the infinite sequence of moments. 
\end{itemize}

Finite size errors have been estimated  comparing the results
on a $64^3$ and a $96^3$ lattice. They turn out to be negligible,
as expected given the sizes of the lattices.
To estimate the statistical errors we performed 
three different simulations using the 50 lowest order moments. 
One
more simulation, this time with the 75 first moments, was done
in order to study the systematic error associated to the truncation
of the sequence of moments. As an example,
Figure~\ref{fig:dos_errors} displays the averaged DOS in the FM phase
$(M_{\mathrm F}=0.5)$ extracted from two different simulations, one
with 50 and the other with 75 moments. 
We see only tiny differences,
which can hardly be appreciated on the scale of the figure.
This small error propagates to the FFE, which is shown in  
Figure~\ref{fig:elibre_errors} for $T=0$ and $\mu=0$. 
There are four sets of points plotted, corresponding to the four 
mentioned simulations.
Again, the differences cannot be appreciated.
The errors in the computed FFE give rise to uncertainties in the 
coefficients $A$'s of Eq.~(\ref{fits}), 
which can be appreciated in Fig.~\ref{fig:coeff_errors}. The largest
errors appear at $\mu=0$ (half-filling). We have also checked that fits to
a polynomial of eight-order do not change the values
of $A_j^{(I)}$ significantly.
\vbox{
\begin{figure}[b!]
\centerline{\epsfig{file=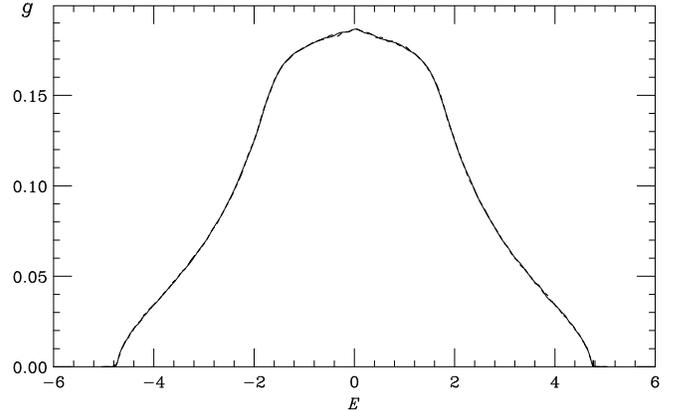,width=55mm,
angle=90}} 
\caption{Averaged DOS  vs. energy for $M_{\mathrm F}=0.5$ on a $64^3$ 
lattice from a
simulation using 50 moments (solid line) and another one
with 75 moments (dashed). The curves can hardly be 
distinguished on this scale.}
\label{fig:dos_errors}
\end{figure}
}

\vbox{
\begin{figure}[t!]
\centerline{\epsfig{file=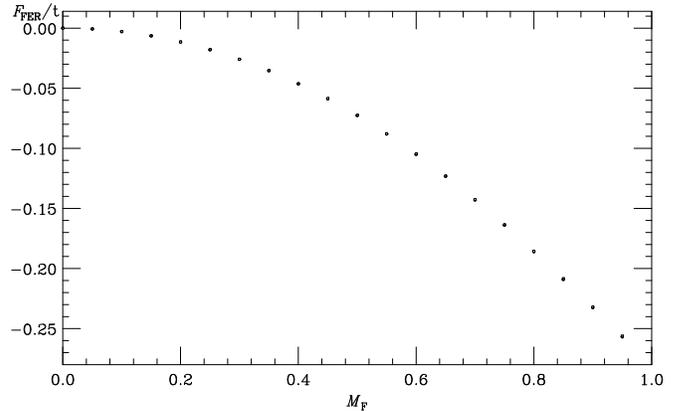,width=55mm,angle=90}} 
\caption{Fermion free energy vs. $M_{\mathrm F}$ from four simulations on
a $64^3$ lattice. Three of them, carried out to estimate the statistical
errors, used 50 moments to reconstruct the
DOS. The other one, aimed to estimate the systematic errors due to
the truncation of the sequence of moments, took into account 75 moments. 
The errors turn out to be so small that cannot be appreciated on the 
scale of the figure.}
\label{fig:elibre_errors}
\end{figure}
}

To summarize, we have checked that the numerical uncertainties inherent
to our numerical approach are well under control and therefore all
the conclusions are robust in this sense. The discrepancies between
our analysis and the true behavior of the system, if they are important, 
must be attributed {\em only} to the Mean-Field {\em ansatz}.

\section{Effect of the Berry phase}

Sometimes it is stated that one can ignore the Berry phase
in the DEM without loosing anything important. 
To investigate this point, let us repeat our analysis of the 
DEM setting the Berry
phase to zero. All we have to do is to compute the DOS of the
hopping matrix ${\cal T}^{\mathrm mod}_{ij}$ of 
Eq.~(\ref{hopmod}). The rest is identical to what we have
discussed in the previous sections.

\vbox{
\begin{figure}[b!]
\centerline{\epsfig{file=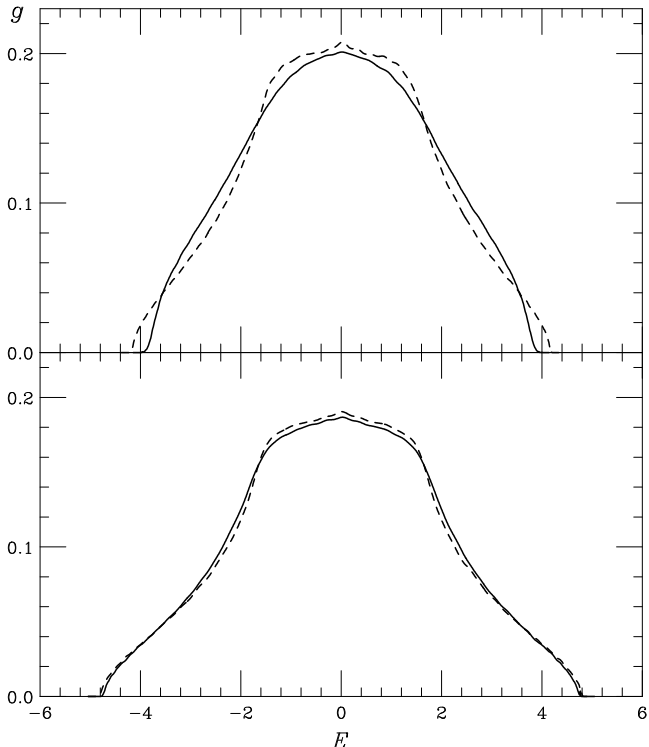,width=100mm,
angle=90}} 
\caption{Averaged DOS in the paramagnetic phase 
($M_{\mathrm F}=0$, upper panel)  and in the ferromagnetic phase
($M_{\mathrm F}=0.5$, lower panel) with (solid)
and without (dashed) Berry phase.}
\label{fig:dos_bp_mod}
\end{figure}
}

Figure~\ref{fig:dos_bp_mod} displays the averaged DOS at
$M_{\mathrm F}=0$ (PM phase) and $M_{\mathrm F}=0.5$ for 
hopping with and without Berry phase. 
At first sight, the differences, although noticeable
(they are much bigger than the errors, 
{\em cf}. Fig.~\ref{fig:dos_errors}), do not
seem very important. However, it happens that the results are
very sensitive to small modifications of the DOS. We shall
see indeed that the presence or absence of the Berry phase is 
crucial for some features of the phase diagram. Thus, the analysis
of errors of the previous section turns out to be extremely
important to give a meaning to our results. Notice that the
effect of the Berry phase is stronger in the disordered phase,
as expected.

\vbox{
\begin{figure}[b!]
\centerline{\epsfig{file=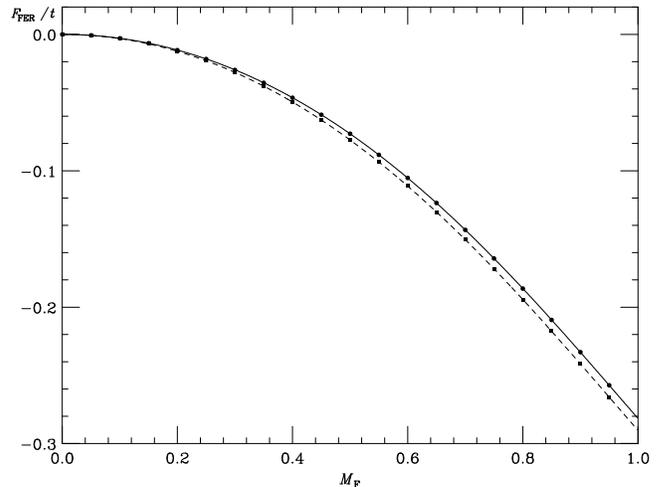,width=65mm,
angle=90}} 
\caption{Fermion free energy at $T=0$ and $\mu=0$
as a function of $M_{\mathrm F}$ with (solid)
and without (dashed) Berry phase. Notice that, in both cases, the free 
energy has been taken vanishing at the origin by convention.}
\label{fig:elibre_bp_mod}
\end{figure}
}

Figure~\ref{fig:elibre_bp_mod} shows the effects of the
Berry phase on the FFE at $T=0$ and $\mu=0$. 
These effects modify the coefficients $A_k^{(I)}$
entering ${\cal F}_{\mathrm Fer}$, which can be
seen in Fig.~\ref{fig:coef_bp_mod}. These coefficients are
very sensitive to the modifications of the DOS
induced by the Berry phase. Of especial relevance
is $A_4^{\mathrm (F)}$, which, as we shall see in the next section,
governs the possibility of having first
order PM-FM transitions. In particular, in the vicinity of $\mu=0$ this
coefficient is negative.
Notice however that without the Berry phase $A_4^{(I)}$ is negative
around $\mu=0$ in a smaller region than with Berry phase,
and it is closer to zero. This fact induces important differences in
the nature of the phase transitions of the model, as we shall see
in the next section.

\section{Phase diagram}

The equilibrium states are determined by the absolute 
minima of the free energy, Eq.~(\ref{entropy}), respect to the order 
parameters, ${\cal M}$. The minima determine the phases 
and the phase
boundaries. Given that we know ${\cal F}$ as a function of 
${\cal M}$, the problem of determining the equilibrium states is 
reduced to numerical minimization of a function of a single
variable. This is indeed the way we proceed. It is however illuminating 
to get some insight by a semi-analytic treatment of the problem.
As we have seen, to a very good approximation, the FFE
is a polynomial of sixth degree in 
${\cal M}$. The entropy (\ref{entro}) can also be expanded in 
powers of ${\cal M}$ around ${\cal M}=0$:

\begin{equation}
{\cal S}_{\mbox{\scriptsize\boldmath$h$}}({\cal M})=
-\left(\frac{3}{2}{\cal M}^2+\frac{9}{20}{\cal M}^4+
\frac{99}{350}{\cal M}^6+\ldots\right) \, .
\label{entro_expanded}
\end{equation}

\noindent
Hence, we find the Landau expansion of the free energy in powers of
the order parameter:

\begin{equation}
{\cal F}({\cal M})=c_2{\cal M}^2+c_4{\cal M}^4+
c_6{\cal M}^6+\ldots \, .
\end{equation}

\vbox{
\begin{figure}[b!]
\centerline{\epsfig{file=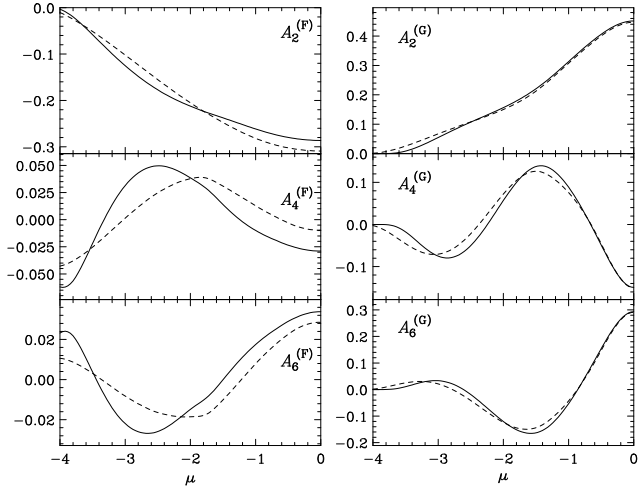,width=65mm,
angle=90}} 
\caption{Coefficients of the fit (\protect\ref{fits})
of the fermion free energy
at $T=0$ as a function of $\mu$ with (solid)
and without (dashed) Berry phase.}
\label{fig:coef_bp_mod}
\end{figure}
}
\noindent
The coefficients of the expansion are

\begin{eqnarray}
c_2&=&\frac{3}{2}T\:+\:N J_{\mathrm AF}\:+\:
A_2^{(I)}(T,\mu)\, ,\\
c_4&=&\frac{9}{20}T\:+\:A_4^{(I)}(T,\mu)\, ,\label{coef_c4}\\
c_6&=&\frac{99}{350}T\:+\:A_6^{(I)}(T,\mu)\, ,
\end{eqnarray}
where $N$ was defined right after Eq.~(\ref{entropy}).

The free energy has the symmetry ${\cal M}\rightarrow -{\cal M}$.
At high $T$, the entropic term dominates and the minimum of
${\cal F}$ is at ${\cal M}=0$. As the temperature decreases,
the internal energy becomes more important and the absolute minimum
of ${\cal F}$ can be located at ${\cal M} \neq 0$. The phase 
transition will be {\em continuous} when the absolute minimum at 
the origin changes to a maximum, i.e.:

\begin{equation}
c_2\;=\;0 \hspace{0.25truecm} \Longrightarrow \hspace{0.25truecm}
\frac{3}{2}T_{\mathrm c}\:+\:NJ_{\mathrm AF}\:+\:
A_2^{(I)}(T_{\mathrm c},\mu) \;=\; 0 
\end{equation}

At a {\em first order transition}
three minima, ${\cal M}=0$ and ${\cal M}=\pm{\cal M}_0\neq 0$,
are degenerate. Three conditions must hold:

\begin{itemize}
\item[i)] Minimum at ${\cal M}=0$:
\begin{equation}
c_2>0\,.
\end{equation}
\item[ii)]Minimum at ${\cal M}_0$:
\begin{equation}
2c_2+4c_4{\cal M}_0^2+6c_6{\cal M}_0^4=0\,.
\end{equation}
\item[iii)]Degeneracy (${\cal F}(0)={\cal F}({\cal M}_0)$):
\begin{equation}
c_2+c_4{\cal M}_0^2+c_6{\cal M}_0^4=0\,.
\end{equation}
\end{itemize}

\vbox{
\begin{figure}[t!]
\centerline{\epsfig{file=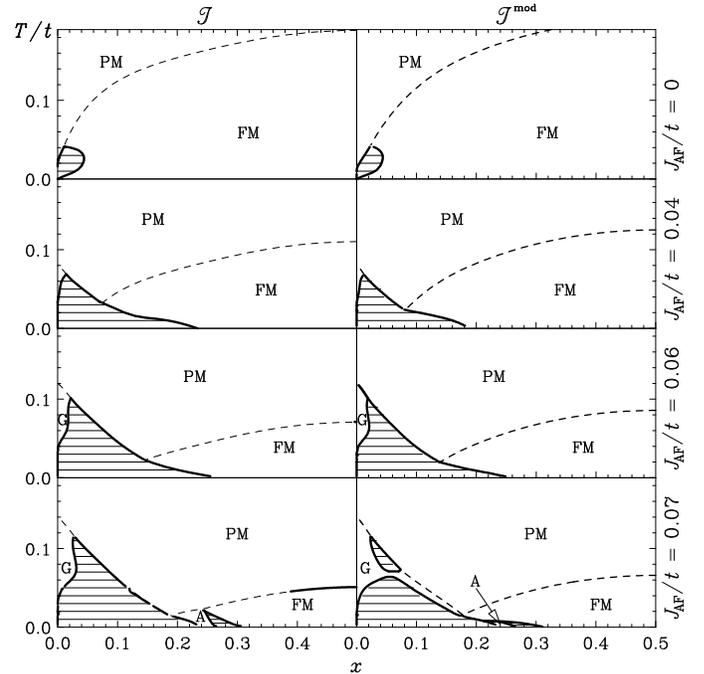,width=90mm,
angle=90}}
\caption{Phase diagram of the DEM 
in the plane $(x,T/t)$, for several values of
$J_{\mathrm AF}/t$. The left (right) part corresponds to the model
with (without) Berry phase. Solid (dashed) lines represent first (second)
order transitions and the zones with stripes are phase separation
regions. The onset for first order
PM-FM transition is at $J_{\mathrm AF}\approx 0.06$ in the model
with Berry phase, while such transitions do not appear if the Berry
phase is neglected.}
\label{fig:Jconst}
\end{figure}
}
\noindent
The solution of these three equations is:

\begin{eqnarray}
& & {\cal M}_0^2=-2c_2/c_4 
\label{sponmag} \\
& & c_6=c_4^2/(4c_2) 
\label{t_crit} \\
& & c_2>0 ; \hspace{0.5truecm}
c_4<0 ; \hspace{0.25truecm} c_6>0
\label{compatib}
\end{eqnarray}

\noindent
Eq.~(\ref{sponmag}) gives the spontaneous magnetization; Eq. 
(\ref{t_crit}) determines the critical temperature $T_{\mathrm c}$ of
the first order transition as a function of $\mu$ and
$J_{\mathrm AF}$; Eq.~(\ref{compatib}) sets necessary conditions 
(real ${\cal M}_0$) for a first order transition to happen. 
We see that to have a first order PM-FM transition 
we must have $A_4^{(I)}(T_{\mathrm c},\mu)<0$. 
Fig.~\ref{fig:coeff_errors} shows
that in particular this is possible around half filling.

The boundaries between first and second order lines are 
tricritical points. They are determined by the conditions:

\begin{eqnarray}
c_2=0 \hspace{0.5truecm} & \Longrightarrow &
\hspace{0.5truecm} J_{\mathrm AF}^{\mathrm t}=-\frac{3}{2N} T_{\mathrm t}-
\frac{A_2^{(I)}(T_{\mathrm t},\mu)}{N} \\
c_4=0 \hspace{0.5truecm} & \Longrightarrow &
\hspace{0.5truecm} T_{\mathrm t}=-\frac{20}{9}A_4^{(I)}
(T_{\mathrm t},\mu)
\hspace*{1.5truecm}
\end{eqnarray}

With these ingredients, we are able to discuss the phase
diagram of the DEM. It has been shown in \cite{Aetal00} that
the phase diagram of double exchange systems is richer than 
previously anticipated and differs substantially from that of 
more conventional itinerant ferromagnets. Moreover, it is consistent 
with the magnetic properties of manganites 
\cite{Tetal97,Uetal99,Fetal99,Fetal96,Fetal98,Mira99,Font99,Metal97,Hwetal95} 
(see section IX).
\vbox{
\begin{figure}[t!]
\centerline{\epsfig{file=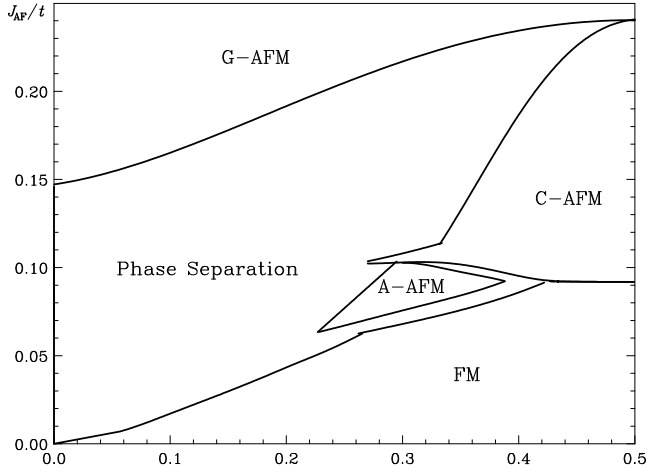,width=65mm,
angle=90}} 
\caption{Phase diagram of the DEM without Berry phase
in the plane \mbox{$(x,J_{\mathrm AF})$} at $T=0$.}
\label{fig:T0}
\end{figure}
}
We shall not repeat here the analysis of the phase diagram of the
DEM carried out in \cite{Aetal00}. Let us concentrate on the effects of
the Berry phase. Figure~\ref{fig:Jconst} displays the phase diagram
in the plane $(x,T/t)$, for several values of $J_{\mathrm AF}$. 
The left part corresponds to the model that includes the
Berry phase and on the right the Berry phase is neglected.
For $J_{\mathrm AF}<0.06$ both phase diagrams are very similar. 
The transition temperature is slightly higher (13\%) at half filling in 
the model that ignores the Berry phase. At low filling, both phase diagrams 
are almost identical. However, for $J_{\mathrm AF}>0.06$ important differences
arise. A first order PM-FM transition develops around half filling if the 
Berry phase is taken into account. The onset for such behavior is 
$J_{\mathrm AF}=0.06$. In contrast, the PM-FM transition around half filling 
remains continuous for any value of $J_{\mathrm AF}$ if
the Berry phase is neglected. The explanation is simple: the coefficient
$A_4^{\mathrm (F)}$, although negative, has a too small absolute value to
drive a first order transition. This negative coefficient would
become effective if the critical temperature were much lower, what can be 
achieved by increasing
the AF superexchange coupling. But in this case the competition between
FM and AF is so strong that the transition at half filling takes
place between PM and A-AFM phases, and it is second order. First order
PM-FM transitions only appear if the Berry phase is properly taken
into account.

At $T=0$, the phase diagrams are similar in both cases, and we only
display that of the model without Berry phase, in Fig.~\ref{fig:T0}.
The discussion of \cite{Aetal00} applies to this case without any
modification.

Let us end this section with the analysis of the phase transitions
at finite applied magnetic field, $B$. The  first order PM-FM transition
around half filling survives under an applied magnetic field. 
In this case, the order 
parameter, $M_{\mathrm F}$, is non-zero in both phases, but suffers
a jump on a line in the plane $(B,T)$. The line
ends at a critical point, $(B^*,T^*)$, which has 
a certain magnetization $M_{\mathrm F}^*$. The critical field
can be measured and is of interest~\cite{Betal99,Amaral00}. 
Let us compute it. The free energy in the presence of a magnetic
field, $B$, is:

\begin{equation}
{\cal F}(M_{\mathrm F})\;=\;c_2M_{\mathrm F}^2\:+\:c_4M_{\mathrm F}^4
\:+\:c_6M_{\mathrm F}^6\:-\:BM_{\mathrm F}\,  .
\end{equation}

\noindent
The magnetic field shifts the three degenerate
minima of the zero-field PM-FM first order 
transition and lifts the degeneracy. By tuning
(increasing) the temperature it is possible to
get two degenerate minima again, 
and a first order transition
takes place. In this way, we get a transition line
in the $(B,T)$ plane. Increasing $B$, the two degenerate
minima become closer. At the critical field, $B^*$,
both minima coalesce at some point $M_{\mathrm F}^*$,
and the transition disappears. When this happens, 
the three first derivatives of ${\cal F}$ respect $M_{\mathrm F}$
vanish, and the fourth is positive. These conditions
read:
 
\begin{eqnarray}
{\cal F}^\prime(M_{\mathrm F}^*)=0 &\Rightarrow&
B^*=2c_2M_{\mathrm F}^*+4c_4M_{\mathrm F}^{*\,3}+6c_6M_{\mathrm F}^{*\,5} \\
{\cal F}^{\prime\prime}(M_{\mathrm F}^*)=0 &\Rightarrow&
2c_2+12c_4M_{\mathrm F}^{*\,2}+30c_6M_{\mathrm F}^{*\,4}=0 \\
{\cal F}^{\prime\prime\prime}(M_{\mathrm F}^*)=0 
&\Rightarrow&
24c_4 M_{\mathrm F}^*+120c_6M_{\mathrm F}^{*\,3}=0 \\
{\cal F}^{(iv)}(M_{\mathrm F}^*)>0 &\Rightarrow&
24c_4+360c_6M_{\mathrm F}^{*\,2}>0.
\end{eqnarray}

\noindent
These equations determine 
$B^*$, $T_{\mathrm c}$ and $M_{\mathrm F}^*$
as a function of $\mu$ (or $x$) and 
$J_{\mathrm AF}$. The critical temperature $T_{\mathrm c}$
varies very little from its value at $B=0$.
Fig.~\ref{fig:hcrit} displays
$B^*$, in units of $10^{-4}\,t$, versus $x$,
for $J_{\mathrm AF}=0.08\,t$. In physical units,
using $t\approx 0.166\,{\mathrm eV}$, $B^*$ varies from 
0.6 Tesla at $x=0.33$ to 2.2 Tesla at $x=0.5$.
Recent measurements in
${\mathrm La_{0.6}\,Y_{0.07}\,Ca_{0.33}\,MnO_3}$ 
gave a critical field of 1.5 Tesla \cite{Amaral00}.

\vbox{
\begin{figure}[t!]
\centerline{\epsfig{file=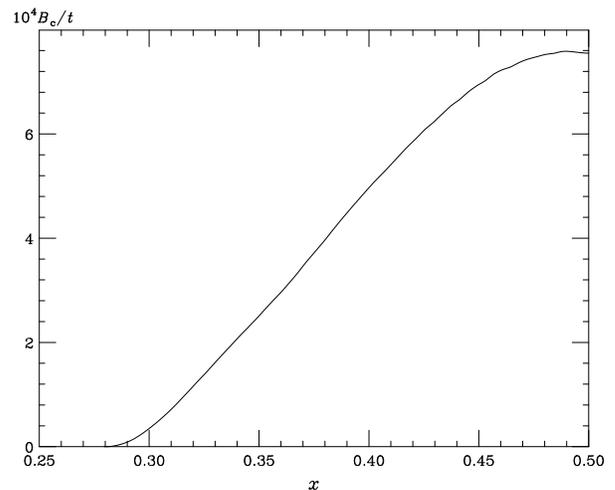,width =65mm,
angle=90}} 
\caption{Critical magnetic field (in units of $10^{-4}\,t$) 
versus $x$ at $J_{\mathrm AF}=0.08\, t$.}
\label{fig:hcrit}
\end{figure}
}
\section{Comparison with other calculations}\label{COMPARISON}
\subsection{Rigid band mean field approximation.}
The main conclusion of the variational Mean-Field
technique applied to the DEM is the prediction of
a first order PM-FM transition at half-filling
and its vicinity for 
\mbox{$J_{\mathrm AF}/t\in [0.06,0.1]$}.
This is in sharp contrast with the widely used
Mean-Field approach devised by de Gennes in 1960 \cite{G60}, 
which predicts a second
order PM-FM at half-filling for any value of 
$J_{\mathrm AF}$. Let us see briefly
what are the differences between these two approaches
that yield different qualitative behavior.

The difficulty in the Mean-Field approach to the DEM
lies in calculating the contribution of the fermions to the
Mean-Field free energy. In
the variational method discussed here, we compute it exactly
through a numerical simulation. As we have already 
mentioned, the only approximation is the 
Mean-Field {\em ansatz} for the Boltzmann weights of the spin configurations.
On the other hand,
de Gennes suggested
that the fermion free energy might be well
approximated by the free energy of an 
assembly of fermions propagating on a crystal 
with an homogeneous hopping parameter given
by the average of the spin-dependent hopping parameter 
over the Mean-Field spin configurations.
The de Gennes' method neglects the influence of the Berry phase.
In this approach.
the electronic DOS depends on the spin configuration only through
the hopping parameter (in this case, without 
Berry phase):
$g(E;\mbox{\boldmath$S$})=
g(E;{\cal T}^{\mathrm mod}(\mbox{\boldmath$S$}_i\cdot
\mbox{\boldmath$S$}_j))$.
In mathematical terms, de Gennes's approximation
is carried out through the following substitution:

\begin{eqnarray}
\left\langle g(E;\left|{\cal T}(\mbox{\boldmath$S$}_i\cdot
\mbox{\boldmath$S$}_j)\right|)
\right\rangle_{\mbox{\scriptsize\boldmath$h$}}&\longrightarrow&
g(E;\langle\left|{\cal T}(\mbox{\boldmath$S$}_i\cdot
\mbox{\boldmath$S$}_j)\right|
\rangle_{\mbox{\scriptsize\boldmath$h$}})\nonumber\\
&&=g_0(E;{\cal T}_0)
\end{eqnarray}

\noindent
where $g_0(E;{\cal T}_0)$ is the DOS of free
fermions with hopping 

\begin{eqnarray}
{\cal T}_0(h)&=&
\langle\left|{\cal T}(\mbox{\boldmath$S$}_i\cdot
\mbox{\boldmath$S$}_j)\right|
\rangle_{\mbox{\scriptsize\boldmath$h$}}\nonumber \\
&=&-\frac{2}{J_{1/2}^2(-{\mathrm i}h)}
\sum_{l=0}^\infty\frac{J_{l+1/2}^2(-{\mathrm i}h)}
{(2l-1)(2l+3)}, 
\label{hop_dg}
\end{eqnarray}

\noindent
and $J_\nu(z)$ is the Bessel function.

Since the hopping is homogeneous, the fermion free energy 
is known analytically. At $T=0$ and half-filling 
($\mu=0$) it is:

\begin{eqnarray}
{\cal F}_{\mathrm Fer}&=&\int_{-{\cal T}_0 W_0}^0
dE\,g(E;{\cal T}_0)E\\
&=&-{\cal T}_0(h)\int_0^{W_0}dE\,g_0(E;1)E
\end{eqnarray}

\noindent
All the dependence in the magnetization is contained in
${\cal T}_0(\mbox{\boldmath$h$})$. The expansion in powers
of the magnetization follows straightforwardly from 
Eqs.~(\ref{hop_dg}) and~(\ref{tanh}). It yields: 

\begin{equation}
\frac{1}{t}{\cal T}_0(h(M_{\mathrm F}))=\frac{2}{3}+
\frac{2}{5}M_{\mathrm F}^2-
\frac{6}{175}M_{\mathrm F}^4-\frac{18}{875}M_{\mathrm F}^6+\ldots
\end{equation}

\noindent
The coefficient of $M_{\mathrm F}^4$ in ${\cal F}_{\mathrm Fer}$ is
{\em positive}. Hence, the PM-FM phase transition at 
half-filling can only be continuous. We have also checked
that this remains true when we keep the contribution
of all powers of $M_{\mathrm F}$ to ${\cal F}_{\mathrm Fer}$.

The fermions in de Gennes's approach propagate only
on perfect crystals. In the truly variational
Mean-Field presented in this work, the fermions
propagate on the disordered spin background generated
by the Mean-Field $\mbox{\boldmath$h$}$. This appears
as an important ingredient that leads the predictions
closer to the phenomenology, as we have shown in 
Ref.~\cite{Aetal00}.

\subsection{Dynamical Mean Field Approximation.}
This method allows for an improvement on the treatment of the 
electronic contribution to the self energy. In the PM phase, the
density of states is proportional
to that in the fully ferromagnetic phase,
like in de Gennes treatment. The only difference is that
the constant of proportionality is $1/\sqrt{2}$ and not $2/3$. 
Below $T_{\mathrm c}$, the density of states is calculated self consistently, 
through a self energy which can be written as:
\begin{equation}
\Sigma \left( E ; \mbox{\boldmath$S$}_i  \right) = 
\langle | {\cal T} ( \mbox{\boldmath$S$}_i \cdot  \mbox{\boldmath$S$}_j) |^2 
g \left( E ; \mbox{\boldmath$S$}_j \right) 
\rangle_{\scriptstyle\mbox{\boldmath$S$}_j }
\end{equation}
and:
\begin{equation}
g ( E ; \mbox{\boldmath$S$}_i ) = \frac{1}{E - \Sigma 
\left( E ; \mbox{\boldmath$S$}_i  \right) }
\end{equation}
Finally, the average 
$\langle \cdots \rangle_{\scriptsize\mbox{\boldmath$S$}_j}$ is carried out 
defining a probability distribution, ${\cal P} (\mbox{\boldmath$S$}_j )$, 
which depends self consistently
on the free energy associated with a site with magnetization
$\mbox{\boldmath$S$}_j$ immersed in the lattice described by 
${\cal P} (\mbox{\boldmath$S$}_j)$.

Our approach is similar to the dynamical mean field approximation, but differs
from it in two aspects:

i) The electronic density of states is calculated in a cubic lattice, instead of
using the semielliptical DOS valid in the Bethe lattice with infinite
coordination.

ii) We use a variational {\em ansatz} for 
${\cal P} ( \mbox{\boldmath$S$}_j)$, 
instead of determining it fully self consistently. 

Point i) allows us to consider effects of the lattice geometry, and the 
influence
of the Berry's phase, as discussed above. 
At zero temperature, where both approaches become exact for their respective 
lattices,
we find phases which can only
be defined in a 3D cubic lattice.

If the transition is continuous, the distribution
${\cal P} (\mbox{\boldmath$S$}_j)$ can be expanded on the deviation from 
the isotropic
one, ${\cal P} (\mbox{\boldmath$S$}_j) = {\mathrm const.}$ ,
 in the PM phase. The {\em ansatz} that we use has the correct behavior 
sufficiently close to $T_{\mathrm c}$, so that both approaches will predict
the same value of $T_{\mathrm c}$, for a given lattice.
One must be more careful in the study of discontinuous transitions. 
Our {\em ansatz}
introduces an approximation in the ordered phase (which disappears at 
$T=0$).
However, near a first order transition we do not expect divergent 
critical fluctuations, so that our approach should give qualitative, and
probably semiquantitative correct results as compared to the DMFA,
in lattices where the latter is exact.
\subsection{Hierarchy of approximations.}

We are tempted to design a hierarchy of approximations,
ordered according to the coefficient $A^{\mathrm (F)}_4$ of the 
$M_{\mathrm F}^4$ term in the Landau expansion of the fermion free
energy, as follows:

\begin{itemize}
\item[1)]
de Gennes's approximation:
$A^{\mathrm (F)}_4>0$ and the PM-FM transition is second order.
\item[2)]
Exact variational computation without Berry phase:
$A^{\mathrm (F)}_4<0$ but $|A^{\mathrm (F)}_4|$ too small to produce first
order PM-FM transitions, see Eq.~(\ref{coef_c4}).
\item[3)]
Exact variational computation with Berry phase:
$A^{\mathrm (F)}_4<0$ and $|A^{\mathrm (F)}_4|$ large enough to produce
first order PM-FM transitions.
\end{itemize}

\section{Conclusions}

We have presented a detailed analysis of the variational
Mean Field technique. This method can be useful in any
situation where non self-interacting fermions are coupled
to classical continuous degrees of freedom. Within this method,
the fermionic contribution to the free energy is calculated exactly,
and, later on, the variational Mean Field method is applied to the
classical degrees of freedom. As an example, we have chosen the
Double Exchange Model, both with and without Berry phase. The
phase diagram has been obtained in both situations.

We have shown that the Berry phase is crucial in order
to get first order PM-FM phase transitions around half filling. 
Such transitions
are second order if the topological effects associated to
the Berry phase are neglected. Thus,  the dimensionality of 
the lattice plays a very important role in the structure
of the phase diagram.

Some earlier Mean-Field computations \cite{G60,AGG99}
approximate the
fermion free energy by that of an assembly of fermions
propagating on a perfect crystal with an homogeneous
hopping parameter averaged over the spin configurations.
They yield second order FM-PM transitions in the vicinity of
half-filling. The propagation of the fermions in the 
disordered spin background generated by the Mean-Field
is another crucial ingredient to get discontinuous 
PM-FM transitions at half-filling. More modern approaches,
as the DMFA~\cite{FURUKAWA} cannot deal
with three dimensional effects as the Berry phase either.

As shown in \cite{Aetal00}, the variational Mean-Field described
in the present work leads to results that are consistent with the 
phenomenology of the magnetic properties of the manganites
${\mathrm La}_{{\mathrm 1}-x}\,{\mathrm (Sr, Ca)}_x\,{\mathrm MnO_3}$, in 
the range $0.3\le x\le 0.5$, in
particular with the fact that for materials with a high transition
temperature, the PM-FM transition is continuous while for those with
low $T_{\mathrm c}$ is not. Moreover, the order of magnitude of our estimate 
of the critical field for which histeretic effects disappear agrees with
the experimental findings in 
${\mathrm La_{0.60}\,Y_{0.07}\,Ca_{0.33}\,MnO_3}$ \cite{Amaral00}.
Also the phase diagram obtained by substitution of a trivalent rare
earth for another one with smaller ionic radius (i.e. compositional
changes that do not modify the doping level) is in remarkable
agreement with our results.

Of course, the DEM itself can also be highly improved. 
For instance, one should include the orbital degeneracy, 
which is known to play an important role, and other
elements like phonons and Jahn-Teller distortions. 
The variational Mean-Field approach can be applied
with the same techniques presented in this work 
whenever the bosonic fields that interact with the
electrons can be treated as classical.

\vspace{0.5truecm}

{\sc Acknowledgements}

We are thankful for helpful conversations to L. Brey, J. Fontcuberta,
G. G\'omez-Santos, C. Simon, J.M. de Teresa, and especially to R.
Ibarra and V. S. Amaral.  V. M.-M. is a MEC fellow.  
We acknowledge financial support
from grants PB96-0875, AEN97-1680, AEN97-1693, AEN97-1708, AEN99-0990
(MEC, Spain) and (07N/0045/98) (C. Madrid).

\appendix
\section{The method of moments}
In this appendix we include, for completeness, some details on the
method of moments~\cite{MOMENTI}. For a complete mathematical background
we refer to Ref.~\cite{CHIHARA}. The method of moments allows to obtain
some statistical properties of large matrices (as the density of
states or the dynamical structure factors, in general quantities
depending on two-legged Green functions), without actually
diagonalizing the matrices.  Regarding the density of states, once one
recognizes that it is a probability function whose moments can be
obtained by iteratively multiplying by ${\cal T}$ the initial random
vector $|v\rangle$, it is clear that the classical Stieljes
techniques~\cite{CHIHARA} can be used. Ought to the fact that the matrix 
${\cal T}$ is sparse, and using the random vector trick,
Eq.~(\ref{RANDOMVECTOR}),
the moments can be calculated with order $V$ operations. Since the
spectrum of the matrix ${\cal T}$ lies between $-6t$ and $6t$ for any
spin configuration, the Stieljes method is guaranteed to converge.
The procedure is as follows: one first introduce the resolvent
\begin{equation}
R(z)=\int_{-6t}^{6t} dE' \,\frac{g(E')}{z-E'}\,,
\end{equation}
that has a cut along the spectrum of  ${\cal T}$, with discontinuity

\begin{equation}
2\pi g(E)= {\cal I}m \lim_{\epsilon\to 0} \left[R(E-{\mathrm i}\epsilon)-
R(E+{\mathrm i}\epsilon)\right]\,.
\label{RECUPERACION}
\end{equation}

The resolvent can be obtained from the orthogonal polynomials of the
$g(E)$, with the monic normalization:
\begin{equation}
P_n(E)\;=\;E^n\:+\:C_{n-1}\,E^{n-1}\:+\:\ldots
\end{equation}
\begin{equation}
\delta_{n,m} \;\propto\; \int_{-6t}^{6t} \, dE\, g(E)\, P_n(E)\, P_m(E)\, ,
\end{equation}
with $P_0=1$, $P_{-1}=0$, and $n,m=0,1,\ldots$ The polynomials verify the 
following recursion relation:
\begin{equation}
P_{n+1}(E)\;=\;(E-a_n)\,P_n(E)\:-\:b_n\,P_{n-1}(E)\, ,
\end{equation}

\noindent
with the coefficients $a_n$ and $b_n$ given by:

\begin{eqnarray}
a_n&=&\frac{\int_{-6t}^{6t}\,dE\,g(E)\, E\, P_n^2(E)}
{\int_{-6t}^{6t}\, dE\, g(E)\, P_n^2(E)}\,,\\
b_n&=&\frac{\int_{-6t}^{6t}\, dE\, g(E)\, P_n^2(E)}
{\int_{-6t}^{6t}\, dE\, g(E)\, P_{n-1}^2(E)}\,.
\end{eqnarray}
The coefficient $b_0$ is arbitrary and 
is conventionally settled to one.

The resolvent has a representation in terms of a continued fraction
as follows:
\begin{equation}
R(z)=\frac{1}{z-a_0-\frac{b_1}{z-a_1-\frac{b_2}{z-a_2-\ldots}}}\,.
\end{equation}
If one truncates the continued fraction, the resolvent would be
approximated by a rationale function, which does not have a cut and
use Eq.~(\ref{RECUPERACION}) is impossible. Fortunately, when,
as in this case, the density of states does not have gaps, the 
coefficients $a_n$ and $b_n$ tends fastly to their asymptotic values 
$a$ and $b$~\cite{CHIHARA}. Thus, one can end the continued 
fraction~\cite{TRUNCACION} with a truncation factor $T(z)$, that verifies
\begin{equation}
T(z)=\frac{b}{z-a-T(z)}\,. 
\end{equation}
Since the previous equation is quadratic in $T(z)$, we find that
$T(z)$ has a branch-cut between $a-2\sqrt{b}$ and $a+2\sqrt{b}$, which
are the limits of the spectrum.

One should not use the moments of the $g(E)$ calculated with 
Eq.~(\ref{RANDOMVECTOR}) to obtain the orthogonal
polynomials (and hence the $\{a_n,b_n\}$), since
this is an extremely unstable numerical procedure. It is better to use 
the recurrence relation:
\begin{equation}
P_{n+1}({\cal T})|v\rangle=({\cal T}-a_n)P_{n}({\cal T})|v\rangle
-b_nP_{n-1}({\cal T})|v\rangle\,,
\end{equation}
\noindent
starting with
\begin{equation}
P_{-1}({\cal T})\,|v\rangle\;=\;0\ ,\;\;\;\;\;\;\;\;  
P_0({\cal T})\,|v\rangle\;=\;|v\rangle\,.
\end{equation}
\noindent
From this, one immediately gets
\begin{eqnarray}
a_n&=&\frac{\langle  P_n({\cal T}) v| {\cal T} P_n({\cal T}) v\rangle}
{\langle  P_n({\cal T}) v| P_n({\cal T}) v\rangle}\,,\\
b_n&=&\frac{\langle  P_n ({\cal T}) v| P_n({\cal T}) v\rangle}
{\langle  P_{n-1}({\cal T}) v| P_{n-1}({\cal T}) v\rangle}\,.
\end{eqnarray} 
In this way, one generates the $N^{\mathrm th}$ orthogonal
polynomial of the matrix (times $v$) recursively, at the price of $N$
multiplications per ${\cal T}$. The cost of this procedure is always of
order $V$ operations. For each random-vector, one first extracts the density of
states through Eq.~(\ref{RECUPERACION}), which is subsequently averaged
over the different $|v\rangle$ and spin realizations.

Let us finally point out that the
above recursion relation is virtually identical to the Lanzcos method 
(the only difference lies on the normalizations). It should therefore not
be pursued for a large number of orthogonal polynomials, without 
reorthogonalization. For the relative modest number of coefficients 
calculated in this work, this has not been needed.


\begin{references}
\bibitem{remark}
We remark that we have chosen these four families of fields for simplicity, 
but this is not a limitation of the method. More exotic ordering as canted
or helicoidal can be also considered.
\bibitem{ACLARACIONMOM} In fact, what one calculates is the 
$50^{\mathrm th}$
orthogonal polynomial (see Appendix), which requires 50
multiplications by the matrix ${\cal T}$, as the calculation of the
$50^{\mathrm th}$ moment would have needed. Extracting the $50^{\mathrm th}$
orthogonal polynomial from the moments without use of the recursion
relations, would have required to calculate 99 moments.  In the text,
the word {\em moment} has been used in this somehow ambiguous way.
\end{references}
\end{document}